\documentclass[amsmath,twocolumn,aps]{revtex4}

\usepackage{longtable}
\usepackage{latexsym}
\usepackage{amssymb}
\usepackage{epsfig}
\usepackage{amsmath}
\usepackage{float}
\usepackage{color}
\usepackage{enumitem}
\usepackage{placeins}
\usepackage[T1]{fontenc}

\newcommand{\be}{\begin{equation}}
\newcommand{\ee}{\end{equation}}
\newcommand{\bea}{\begin{eqnarray}}
\newcommand{\eea}{\end{eqnarray}}

\newcommand{\Tb}{\bar{T}_b}
\newcommand*{\dittostraight}{---\textquotedbl---} 

\begin{document}

\title{ Synergistic tests of inflation } 

\author{
Alkistis Pourtsidou\footnote{E-mail: alkistis.pourtsidou@port.ac.uk}
}

\affiliation{
Institute of Cosmology \& Gravitation, University of Portsmouth, Dennis Sciama Building, Burnaby Road, Portsmouth, PO1 3FX, United Kingdom
}

\begin{abstract}
We investigate the possibility of utilising 21cm intensity mapping, optical galaxy, and Cosmic Microwave Background (CMB) surveys to constrain the power spectrum of primordial fluctuations predicted by single-field slow-roll inflation models. 
Implementing a Fisher forecast analysis, we derive constraints on the spectral tilt parameter $n_{\rm s}$ and its first and second runnings $(\alpha_{\rm s},\beta_{\rm s})$. We show that 21cm intensity mapping surveys with instruments like the Square Kilometre Array, CHIME, and HIRAX, can be powerful probes of the primordial features. 
We combine our forecasts with the ones derived for a COrE-like
CMB survey, as well as for a Stage IV optical galaxy survey similar to Euclid. 
The synergies between different surveys can be exploited to rule out a large fraction of the available inflationary models.
\end{abstract}

\maketitle

\section{Introduction}

Inflation is a period of accelerated expansion in the very early Universe, and it is currently the most compelling candidate theory in order to explain the origin of structure in the cosmos  (for a review, see \cite{Baumann:2009ds, Senatore:2016aui} and references therein). Vanilla inflation generally predicts a homogenous, isotropic, and spatially flat Universe with nearly scale invariant primordial power spectrum and nearly gaussian density fluctuations.    

To be more specific, let us assume a perturbed FRW Universe and denote the scalar perturbations power spectrum as $P_\zeta(k)$. We can define the dimensionless power spectrum as
\be
P_{\rm s}(k) \equiv \frac{k^3}{2\pi^2} P_\zeta(k) \, .
\ee Then we can write
\be
P_{\rm s}(k) = A_{\rm s} \left(\frac{k}{k_\star}\right)^{n_{\rm s}-1+\frac{1}{2}\alpha_{\rm s} {\rm ln}(k/k_\star)+
\frac{1}{6}\beta_{\rm s} {\rm ln}^2(k/k_\star)} \, .
\ee Here $A_{\rm s}$ is the amplitude of the scalar perturbations and $k_\star$ is the pivot scale where the spectral index $n_{\rm s} \equiv {\rm d ln}P_s/{\rm d ln}k$ and its runnings are defined and measured. The spectral index $n_{\rm s}$ is measured by Planck to be close, but not equal, to unity: $n_{\rm s}=0.968 \pm 0.006$ \cite{Ade:2015lrj}. The first running is defined as $\alpha_{\rm s} \equiv {\rm d} n_{\rm s}/{\rm d ln}k$, and its Planck measurement is consistent with zero: $\alpha_{\rm s} = -0.003 \pm 0.007$ \cite{Ade:2015lrj}. Note that a (tentative) non-zero positive second running $\beta_{\rm s} \equiv {\rm d}\alpha_{\rm s}/{\rm d ln}k$ was found in the analysis of \cite{Cabass:2016ldu}: $\beta_{\rm s} = 0.027 \pm 0.013$. The pivot scale for these measurements is $k_\star = 0.05 \, {\rm Mpc}^{-1}$.

In the absence of evidence for non-minimal extensions of the inflationary scenario (we have not observed primordial non-Gaussianities or isocurvature perturbations, for example), single-field slow-roll models are generally favoured. Unfortunately this means that there is a plethora of allowed models and finding the most favoured one requires high precision new data and advanced statistical methods \cite{Martin:2013tda}.

In the simplest, single-field slow-roll inflationary models, the inflaton field that drives inflation is a canonical scalar field $\phi$. The inflaton potential $V(\phi)$ and its derivatives can be directly related to $A_{\rm s}$, $n_{\rm s}$, and its runnings, as well as to the amplitude and index of the tensor perturbations. Defining the slow-roll parameters (evaluated at the field value $\phi_{\star}$ where the pivot scale $k_\star$ exits the Hubble radius during inflation)
\bea \nonumber
\epsilon &=& \frac{M^2_{\rm pl}}{2} \left(\frac{V^\prime}{V}\right)^2 \\ \nonumber
\eta &=& M^2_{\rm pl} \left(\frac{V^{\prime \prime}}{V}\right) \\
\xi &=& M^4_{\rm pl} \left(\frac{V^\prime V^{\prime \prime \prime}}{V^2}\right) \, ,
\eea where $M_{\rm pl}$ is the reduced Planck mass and a prime denotes differentiation with respect to the field $\phi$, we get
\bea \nonumber
1-n_{\rm s} &=& 2\eta - 6\epsilon + ... \\
\alpha_{\rm s} &=& -2\xi  + 16\eta \epsilon -24 \epsilon^2 + ... \, ,
\eea therefore $n_{\rm s}-1$ is first order in slow-roll parameters, $\alpha_{\rm s}$ second order and similarly $\beta_{\rm s}$  is third order (see \cite{Easther:2006qu, Liddle:1994dx} for details). During slow roll, these parameters are very small, $\epsilon \ll 1$ and $|\eta| \ll 1$.
The above formalism gives a general prediction for the size and hierarchy of the runnings. That is, $|\alpha_{\rm s}| \sim 0.001$ and $|\beta_{\rm s}| \sim 10^{-5}$. Any significant deviation from these values would disfavour 
single-field slow-roll inflation. 

In this work we are going to use the Fisher matrix approach to forecast how well future 21cm intensity mapping (IM), optical galaxy, and CMB surveys, can constrain the spectral index and its runnings. CMB temperature and polarization measurements probe the primordial power spectrum $P_s(k)$ and constrain the various quantities it depends on. In our CMB forecasts we will constrain the six $\Lambda$CDM parameters, namely the baryon and cold dark matter densities $w_{\rm b} \equiv \Omega_{\rm b} h^2$, $w_{\rm c} \equiv \Omega_{\rm c} h^2$, the scalar amplitude $A_{\rm s}$, the optical depth to reionization $\tau$, the Hubble parameter $H_0$, and the tilt $n_{\rm s}$, together with the first and second runnings $(\alpha_{\rm s},\beta_{\rm s})$. 
Large scale structure surveys use biased tracers of matter -- for example galaxies or neutral hydrogen (HI) -- to probe the matter power spectrum
\be
P(k) = T^2(k) P_\zeta(k) \, ,
\ee  where $T(k)$ is the transfer function. 
In our large scale structure (LSS) forecasts we will only vary the inflationary parameters $(n_{\rm s},\alpha_{\rm s},\beta_{\rm s})$, considering the rest of the parameters fixed (measured) by the CMB. The same approach was followed in \cite{Munoz:2016owz} for optical and HI galaxy surveys.

The paper is organised as follows: In Section~\ref{sec:surveys} we describe the range of 21cm intensity mapping, CMB, and optical galaxy surveys we are going to use in our forecasts. In Section~\ref{sec:results} 
we describe our formalism for the different types of surveys, derive the Fisher matrix constraints, and discuss the results. The results are also summarised in Tables~\ref{tab:core}, ~\ref{tab:IM}, and~\ref{tab:gal}. In order to assess the synergistic power of future CMB and LSS surveys, we combine our forecasts in Table~\ref{tab:comb}. We conclude in Section~\ref{sec:conclusions}.
Our fiducial cosmology is the Planck 2015 best-fit $\Lambda$CDM model \cite{Ade:2015xua}, with $\alpha_{\rm s} = \beta_{\rm s} = 0$. We also take the tensor-to-scalar ratio $r=0$ for simplicity, since it does not affect the runnings.

\section{The surveys}
\label{sec:surveys}

\subsection{Stage 4 CMB survey}

We consider a future CMB survey with characteristics similar to the proposed COrE satellite mission \cite{DiValentino:2016foa}. We will use the forecasted measurements for the temperature (T), polarization (E), and cross temperature-polarization angular power spectra (see Section~\ref{sec:results} for the relevant formulae). 
The instrument's TT noise power spectrum  is given by
\be
N^{\rm TT}_\ell  = \Delta_T^2 {\rm exp}[\ell(\ell+1)\theta_{\rm FWHM}^2/8{\rm ln}2] \, ,
\ee where $\theta_{\rm FWHM}$ is the full width half maximum of the beam and $\Delta_T$ the sensitivity. 
The EE noise power spectrum is taken to be
\be
N^{\rm EE}_\ell  = 2N^{\rm TT}_\ell \, .
\ee
We assume that the mission covers a fraction of the sky $f_{\rm sky} = 0.7$ with sensitivity 
$\Delta_T = 5 \, {\rm \mu K - arcmin}$ and $\theta_{\rm FWHM} = 4 \, {\rm arcmin}$. We will consider the range of multipoles $\ell_{\rm min} = 10$ to $\ell_{\rm max} = 5000$ in our forecasts, with $k_{\star} = 0.05 \, {\rm Mpc}^{-1}$.

\subsection{21cm intensity mapping surveys}

21cm intensity mapping \cite{Battye:2004re,Chang:2007xk,Seo:2009fq,Ansari:2011bv,Battye:2012tg,Switzer:2013ewa,Fonseca:2016qqw} is a technique that uses HI as a dark matter tracer in order to map the 3D large-scale structure of the Universe. It measures the intensity of the redshifted 21cm line, hence it does not need to detect galaxies but treats the 21cm sky as a diffuse background. This means that intensity mapping surveys can scan large volumes of the sky very fast. They also have excellent redshift information, and can perform high precision clustering measurements \cite{Bull:2014rha, Pourtsidou:2015mia}. 

A radio telescope array similar to the Square Kilometre Array (SKA)\footnote{www.skatelescope.org} performing an intensity mapping survey can be used as an interferometer or as a collection of single dishes, measuring the cross- or auto-correlation signal, respectively. The advantages of one mode of operation over the other are depending on what are the specifications and science goals of the experiment \cite{Bull:2014rha}. In general, if sufficient sky area is scanned the single-dish mode can probe cosmological scales at low redshifts, so it is ideal for late-time Baryon Acoustic Oscillations studies \cite{Bull:2014rha}. However, it has limited angular resolution. An SKA-like interferometer, on the other hand, has very good angular resolution set by its maximum baseline, but due to its small field-of-view limited by the primary beam size it cannot probe large scales unless it operates at low frequencies (high redshifts). Nevertheless, an interferometer with smaller dishes and a high covering fraction (i.e. a compact array) can probe larger scales and has increased sensitivity, especially if it can achieve a large instantaneous field-of-view (FOV) \cite{Pourtsidou:2013hea, Bull:2014rha}. 

The importance of the above characteristics will become evident later on in our analysis. In the following subsections, we will describe the noise properties of an intensity mapping survey using radio arrays operating in single-dish and interferometer mode. We will also catalogue the specific instruments and surveys we are going to use in our forecasts.

\subsubsection{Single-dish mode}

The single-dish noise properties have been described in detail in \cite{Battye:2012tg, Bull:2014rha, Pourtsidou:2015mia}, but we will repeat the analysis here for completeness. The instrument response 
due to the finite angular resolution can be modelled as
\be
W^2 = {\rm exp}\left[-k^2_\perp r(z)^2\left(\frac{\theta_{\rm FWHM}}{\sqrt{8{\rm ln}2}}\right)^2\right] \, ,
\label{eq:response}
\ee where ${\mathbf k}_\perp$ is the transverse wavevector,  $r(z)$ is the comoving distance at redshift $z$ and $\theta_{\rm FWHM} \sim \lambda/D_{\rm dish}$ the beam full width at half maximum of a single dish with diameter $D_{\rm dish}$ at observation wavelength $\lambda = 21(1+z) \, {\rm cm}$. We have ignored the response function in the radial direction $(k_\parallel)$ as the frequency resolution of intensity mapping surveys is very good (of the order of tens of ${\rm kHz}$). 
Considering a redshift bin with limits $z_{\rm min}$ and $z_{\rm max}$, the survey volume $V_{\rm sur}$ will be given by
\be
V_{\rm sur}=S_{\rm area} \int_{z_{\rm min}}^{z_{\rm max}} dz \frac{dV}{dz d\Omega} =  S_{\rm area}\int_{z_{\rm min}}^{z_{\rm max}} dz \frac{c r(z)^2}{H(z)} \, ,
\label{eq:volume}
\ee with $S_{\rm area}$ the sky area the survey scans (in steradians). The pixel volume $V_{\rm pix}$ is also calculated from Eq.~(\ref{eq:volume}), but with pixel area
$
 \Omega_{\rm pix} \simeq 1.13\theta^2_{\rm FWHM}
$ assuming a Gaussian beam, and the corresponding pixel $z$-limits corresponding to the channel width $\Delta f$. Finally, the pixel thermal noise $\sigma_{\rm pix}$ is given by
\be
\sigma_{\rm pix}=\frac{T_{\rm sys}}{\sqrt{\Delta f \, t_{\rm total}(\Omega_{\rm pix}/S_{\rm area})N_{\rm dishes} N_{\rm beams}}} \, ,
\ee with $N_{\rm dishes}$ the number of dishes, $N_{\rm beams}$ the number of beams (feeds) and $t_{\rm total}$ the total observing time, with the combination $t_{\rm total}(\Omega_{\rm pix}/S_{\rm area})$ representing the time spent at each pointing. The system temperature $T_{\rm sys}$ is found by summing the instrument temperature $T_{\rm inst}$ and the sky temperature -- dominated by the galactic synchrotron emission --  $T_{\rm sky} \approx 60 \, (300 {\rm MHz}/\nu)^{2.55} \, {\rm K}$, where $\nu$ the frequency of observation. Note that $T_{\rm sky}$ is usually subdominant to $T_{\rm sys}$ at low redshifts.
In intensity mapping experiments the shot noise can be neglected and the dominant noise contribution comes from the thermal noise of the instrument. The noise power spectrum is then given by
\be
\label{eq:PNsd}
P^{\rm N} = \sigma^2_{\rm pix}V_{\rm pix}W^{-2} \, .
\ee
We are going to consider such a survey using the SKA \cite{Santos:2015bsa}. 

\subsubsection*{{\bf SKA}}

For SKA Phase 1 (SKA1) we are going to set $N_{\rm dishes} = 194$ (that is 130 SKA1-MID dishes and 64 MeerKAT dishes), with $D_{\rm dish} \simeq 15 \, {\rm m}$. We will use $f_{\rm sky}=0.7$ for our presented forecasts, and $t_{\rm total} = 5,000 / 10,000 \, {\rm hrs}$. The redshift range is 
$0.35<z<3.05$ (Band 1) and the instrument temperature is taken to be $T_{\rm inst} = 25 \, {\rm K}$. For the more futuristic SKA2-MID scenario we are going to assume an array with an order of magnitude higher sensitivity. We set the largest scale the array can probe when operating in single-dish mode using $k_{\rm min} \simeq 2\pi (V_{\rm bin})^{-1/3}$, the limit set by the survey (bin) volume. 

\subsubsection{Interferometer mode}

The noise power spectrum for a dual polarization interferometer array assuming uniform antennae distribution is \cite{Zaldarriaga:2003du, Tegmark:2008au}
\be
P^{\rm N} = T^2_{\rm sys}r^2y_\nu\left(\frac{\lambda^4}{A^2_{\rm e}}\right)
\frac{1}{2n(u)t_{\rm total}} \left(\frac{S_{\rm area}}{\rm FOV}\right) \, . 
\ee 
Here, $A_{\rm e}$ is the effective beam area, $r$ is the comoving distance to the observation redshift $z$, and $y_\nu = c(1+z)^2/(f_0 H(z))$ with $f_0 = 1420 \, {\rm MHz}$, the HI rest frame frequency. 
The distribution function of the antennae $n(u)$ is approximated as $n(u)\simeq N^2_f/2\pi u^2_{\rm max}$ for the uniform case, where $N_f$ is the number of elements of the interferometer and $u_{\rm max}\simeq D_{\rm max}/\lambda$ with $D_{\rm max}$ the maximum baseline. 

We are going to consider such a survey using CHIME \cite{Newburgh:2014toa,Bandura:2014gwa}, HIRAX \cite{Newburgh:2016mwi}, and the SKA. 

\subsubsection*{{\bf CHIME}}

CHIME (The Canadian Hydrogen Intensity Mapping Experiment) is a dark energy experiment designed to perform a 21cm intensity mapping survey in the redshift range $0.8<z<2.5$ in order to detect BAOs and constrain dark energy. It is a cylindrical interferometer, consisting of $N_{\rm cyl}= 5$ cylinders ($W_{\rm cyl}= 20 \, {\rm m} \,  \times \,  L_{\rm cyl}= 100 \, {\rm m}$) with $N_{\rm f}= 1024 \, {\rm feeds}$.  The system temperature is taken to be $T_{\rm sys} = 50 \, {\rm K}$ and the maximum baseline $D_{\rm max}=128 \, {\rm m}$. To calculate the noise power spectrum for CHIME, we need to make some approximations in order to model the primary beam, which is anisotropic. Following the approach described in \cite{Bull:2014rha}, the effective area per feed is calculated as
$A_{\rm e} = \eta L_{\rm cyl}W_{\rm cyl}N_{\rm cyl}/N_{\rm f}$, with the efficiency $\eta = 0.7$ and the approximate (isotropic) field-of-view is ${\rm FOV} = (1.22 \lambda / W_{\rm cyl} ) \times 90(\pi/180)$. The largest scale the array can probe is set by $k_{\rm min} = 2\pi W_{\rm cyl}/(r\lambda)$, and the smallest is $k_{\rm max} =  2\pi D_{\rm max}/(r\lambda)$. The sky area for CHIME is $S_{\rm area} = 25,000 \, {\rm deg}^2$ with a total observation time $t_{\rm total} = 10,000 \, {\rm hrs}$.

\subsubsection*{{\bf HIRAX}}

HIRAX is another compact radio interferometer, located in South Africa, which is designed to perform a 21cm intensity mapping survey in the redshift range $0.8<z<2.5$.  HIRAX aims to provide LSS measurements in order to probe dark energy. It will also look for radio transients and pulsars. HIRAX consists of $N_f = 1024$, $6 \, {\rm m}$ dishes, closely packed together in an area with $D_{\rm max} \sim 250 \, {\rm m}$. The sky area for HIRAX is $S_{\rm area} = 15,000 \, {\rm deg}^2$ with a total observation time $t_{\rm total} = 10,000 \, {\rm hrs}$. The system temperature is taken to be $T_{\rm sys} = 50 \, {\rm K}$. HIRAX and CHIME are very complementary (similar science goals, same redshift range, different sky (North / South)).

\subsubsection*{{\bf SKA}}

SKA-LOW is an interferometer that will map the 21cm sky at redshifts $z=6-25$ in order to probe the Epoch of Reionization and the Cosmic Dawn \cite{Pritchard:2015fia}. It can also provide 21cm intensity maps at the post-reionization redshifts  $3<z<5$.
We will consider a futuristic SKA2-LOW-like intensity mapping survey, covering the redshift range $3<z<5$. This has $N_f = 7,000$, $6 \, {\rm m}$ dishes, closely packed together in an area with $D_{\rm max} \sim 700 \, {\rm m}$. We use  $f_{\rm sky} = 0.5$ with a total observation time $t_{\rm total} = 10,000 \, {\rm hrs}$. The instrument temperature is taken to be $T_{\rm inst} = 15 \, {\rm K}$, hence the sky temperature dominates at all redshifts. We are not going to consider SKA1-MID in interferometer mode, as its current design is sparse and its dishes are big.  

\subsection{Stage 4 spectroscopic galaxy survey}

The possibility of constraining the inflationary parameters $(n_{\rm s},\alpha_{\rm s},\beta_{\rm s})$ with galaxy redshift surveys has been investigated in the past (see, for example, \cite{Takada:2005si,  Adshead:2010mc, Huang:2012mr, Chen:2016vvw, Ballardini:2016hpi, Munoz:2016owz}). 
In this work, we will consider a Stage IV spectroscopic optical galaxy survey similar to the forthcoming Euclid satellite mission \cite{Amendola:2016saw}. The survey operates in the redshift range $0.7<z<2$ detecting tens of millions of galaxies in a sky area $S_{\rm area}=15,000 \, {\rm deg}^2$.  In our forecasts for such a survey we will use the number density of galaxies $\bar{n}$ and the galaxy bias $b_g$ given in \cite{Majerotto:2015bra}, where the predicted redshift distribution has been split into 14 bins with $\Delta z=0.1$. 
 
\section{Formalism and Results}
\label{sec:results}

\subsection{CMB}

The CMB power spectra are given by
\be
C^{XY}_\ell = (4\pi)^2 \int dk k^2 T^X_\ell(k) T^Y_\ell(k) P_\zeta(k) \, .
\ee As we have already stated, we are going to use the (unlensed) temperature and E-mode polarization information in our forecasts, so that $\{ X,Y \} = \{ T, E \}$ and $T^X$ are the corresponding transfer functions that do not depend on inflationary parameters. The covariance matrix 
$\mathbf{C}_\ell$ is then given by 
\(
\begin{bmatrix}
    \hat{C}_\ell^{\rm TT}  & C_\ell^{\rm TE}  \\
    C_\ell^{\rm TE}   & \hat{C}_\ell^{\rm EE} 
\end{bmatrix}
\)
where $\hat{C}_\ell = C_\ell + N_\ell$. 

Then the Fisher matrix for a set of parameters $\{ p_i \}$ is given by
\be
\label{eq:Fisher}
F_{ij} = \sum_\ell \frac{2\ell+1}{2}f_{\rm sky}{\rm Tr}\left(\mathbf{C}^{-1}_\ell
 \frac{\partial \mathbf{C}_\ell}{\partial p_i}  \mathbf{C}^{-1}_\ell  \frac{\partial \mathbf{C}_\ell}{\partial p_j} \right) \, ,
\ee 
and the marginalised $1-\sigma$ error on a parameter is $\sigma(p_i) = \sqrt{(F^{-1})_{ii}} $.
Performing the analysis assuming a COrE-like satellite with specifications given in Section~\ref{sec:surveys} we find the $1-\sigma$ uncertainties quoted in Table~\ref{tab:core}. 
We also show the correlation coefficient $r$ for the second running $\beta_{\rm s}$ and a parameter $p$, namely 
\be
r(\beta_{\rm s},p) = \frac{(F^{-1})_{\beta_{\rm s} p}}{\sqrt{(F^{-1})_{\beta_{\rm s} \beta_{\rm s}}(F^{-1})_{pp}}} \, .
\ee
Note that these numbers change depending on the choice of pivot scale (we will discuss this further later on) and the type of measurements and/or priors one employs; in general, there are degeneracies between cosmological parameters and the runnings, and between the runnings themselves.

\begin{center}
\begin{table*}
\begin{tabular}{| l | c | c | c | c | c | c | c | c  |  }
\hline
Model & $\sigma\left(w_{\rm b}\right)$ & $\sigma\left(w_{\rm c}\right)$ & $\sigma\left(A_{\rm s}\right)$ & $\sigma\left(\tau\right)$ & $\sigma\left(H_0\right)$ & $\sigma\left(n_{\rm s}\right)$ &  $\sigma\left(\alpha_{\rm s}\right)$&  $\sigma\left(\beta_{\rm s}\right)$   \\             
\hline
\hline
$\Lambda$CDM+$\alpha_{\rm s}$ & $4.5 \times 10^{-5}$ & $6.5 \times 10^{-4}$ & $1.3 \times 10^{-11}$ & $0.003$ & $0.26$ & $0.0019$ & $0.0025$ & $-$ \\
$\Lambda$CDM+$\alpha_{\rm s}$+$\beta_{\rm s}$ & $4.7 \times 10^{-5}$ & $7.3 \times 10^{-4}$ & $1.5 \times 10^{-11}$ & $0.003$ & $0.29$ & $0.0030$ & $0.0026$ & $0.0058$ \\
\hline
\hline
$r(\beta_{\rm s},p)$ & $-0.27$ & $0.45$ & $0.42$ & $0.23$ & $-0.46$ & $-0.76$ & $-0.15$ & $1$ \\		
\hline	
\end{tabular}
\caption{$1-\sigma$ forecasts for the COrE-like CMB experiment and two models, one including $\Lambda$CDM and the first running $\alpha_{\rm s}$ only, and one adding the second running $\beta_{\rm s}$. We also show the $\beta_{\rm s}$ correlation coefficients.}
\label{tab:core}
\end{table*}
\end{center}

Measurements of $\sigma (\alpha_{\rm s}) \simeq 0.0025$ are not enough to detect the prediction $|\alpha_{\rm s}| \sim 0.001 $, but they can be useful in order to test for significant deviations from the single-field slow-roll scenarios; similar conclusions were drawn in a very recent study \cite{Munoz:2016owz}, which used the proposed ground-based CMB-S4 experiment \cite{Abazajian:2016yjj} to forecast constraints on the same cosmological and inflationary parameters.  Measurements of  $\sigma (\beta_{\rm s}) \sim 0.006$ will be able to confirm or discard the indication for a large, positive $\beta_{\rm s}$ \cite{Cabass:2016ldu}.  
In fact, studies have shown that the constraining power of future CMB missions like the one we have considered here is immense:  using Bayesian analysis and in particular the Jeffreys' scale method, \cite{Martin:2014rqa} found that the number of models that can be ruled out with high statistical significance increases from one third for Planck to three quarters for a Stage 4 CMB mission. This is a notable improvement, and one should also keep in mind that the single-field slow-roll models represent the most pessimistic, minimal scenario (most difficult to constrain). For example, if the indication for a positive $\beta_{\rm s} > \alpha_{\rm s}$ found in \cite{Cabass:2016ldu} is confirmed, we will need to start looking at extended models of inflation \cite{vandeBruck:2016rfv}. 

The prospects of future CMB surveys to probe the inflationary Universe are excellent. It is also important to explore how additional datasets from large scale structure surveys can boost their constraining and discriminating power even further. Motivated by this potential synergy between CMB and LSS surveys, we will now move on to investigate how 21cm intensity mapping surveys performed in a wide range of post-reionization redshifts can be used to place constraints on the scalar spectral index and its runnings.

\subsection{Intensity Mapping}

The mean 21cm emission brightness temperature is given by (see \cite{Battye:2012tg} for a detailed derivation)
\be
\Tb(z) = 180 \Omega_{\rm HI}(z)h\frac{(1+z)^2}{H(z)/H_0} \, {\rm mK} \, ,
\ee where $\Omega_{\rm HI}$ is the HI density and $H_0\equiv 100h$ is the value of the Hubble rate $H(z)$ today. 

Neglecting --for the moment-- redshift space distortions (RSDs), we can model  the HI power spectrum as
\be
 P^{\rm HI}(k,z)=\Tb^2 b^2_{\rm HI} P(k,z) \, ,
\ee where $P$ is the matter power spectrum and $b_{\rm HI}$ the HI bias, assumed to be scale independent and deterministic on linear scales. 
In our forecasts we will consider $\Omega_{\rm HI}$ and $b_{\rm HI}$ known.
Currently, these factors are poorly constrained \cite{Padmanabhan:2014zma}. However, studies have shown that forthcoming intensity mapping surveys using the SKA and its pathfinders (for example MeerKAT), as well as cross-correlations with optical galaxy surveys, will be able to place stringent constraints on these parameters across a wide range of redshifts \cite{Pourtsidou:2016dzn}. Constraints will also come from galaxy surveys and damped Lyman-$\alpha$ system measurements, in combination with results from simulations and theoretical modelling \cite{Padmanabhan:2014zma}. For our fiducial models of the HI density, bias, and $\Tb$, we use the fits from \cite{Bull:2014rha}.

\begin{figure}[H]
\centering
\includegraphics[width=\columnwidth]{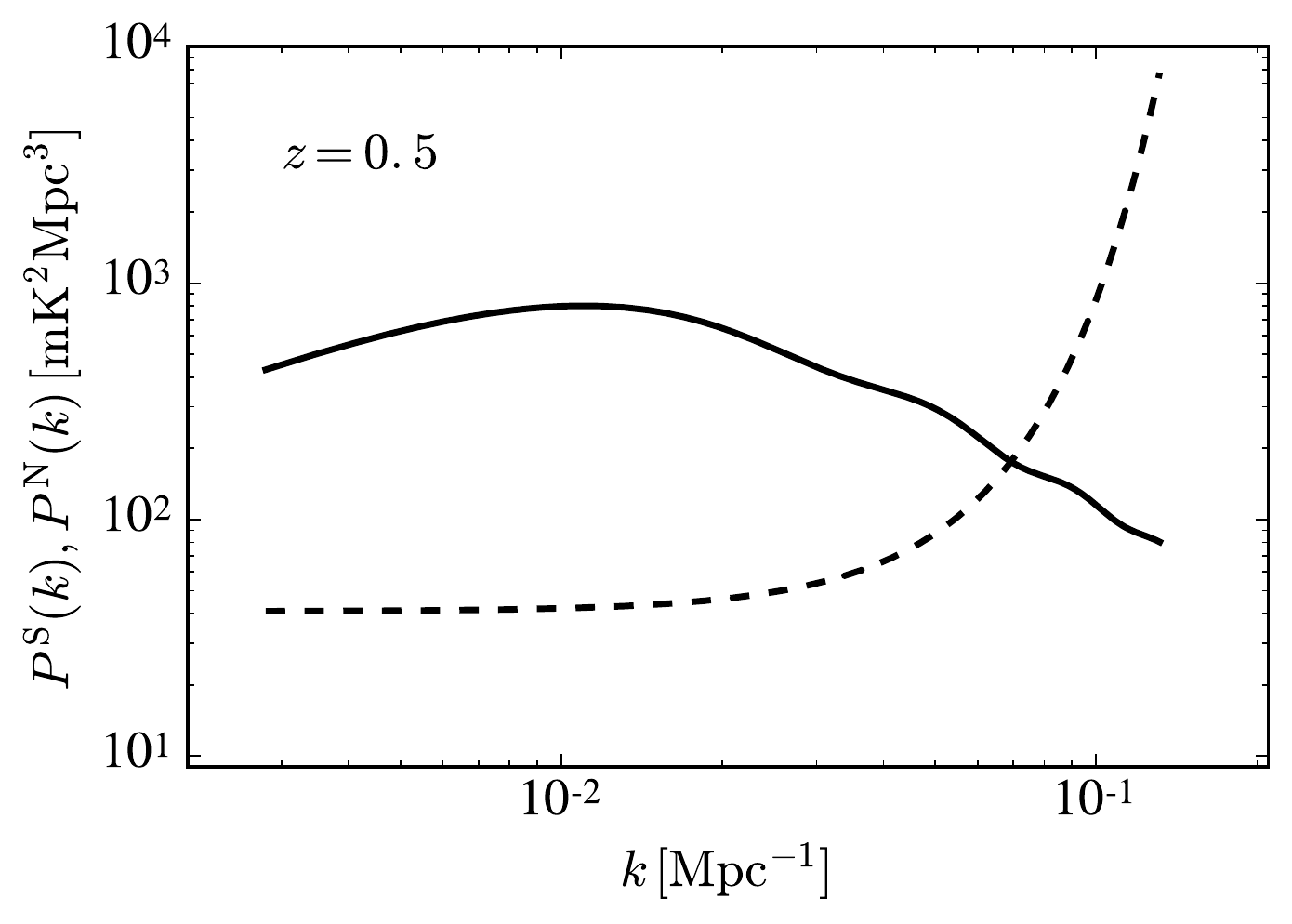}
\caption{The HI (solid) and thermal noise (dashed) power spectra at $z=0.5$ for the chosen SKA1-MID IM survey parameters (see main text for details).}
\label{fig:sd-PN}
\end{figure}

In Figure~\ref{fig:sd-PN} we plot the HI and thermal noise power spectra for an SKA1-MID array and $f_{\rm sky} \sim 0.7$, $t_{\rm total} = 5,000 \, {\rm hours}$, at a redshift bin centred at $z=0.5$ with width $\Delta z = 0.1$. We note that for this plot we have used a simplified response function by setting $k_\perp \sim k$ in Eq.~(\ref{eq:response}). In our forecasts below 
we will include the redshift space distortions contribution in the HI signal and implement the full (anisotropic) modelling of the instrument's response. The single-dish mode is useful for observing large and ultra-large scales assuming sufficient sky area is scanned, while the noise diverges quickly as we reach the limits set by the beam
resolution. Since the beam resolution decreases with redshift, with the single-dish mode we lose the advantage of using the smaller -- but still linear -- transverse scales at higher redshifts.

Including RSDs, the HI signal power spectrum in redshift space can be written as 
\be
P^{\rm S} \equiv P^{\rm HI}(k,z;\mu)=\Tb^2  [b^2_{\rm HI}+f\mu^2]^2P(k,z) \, ,
\ee  
where $\mu = \hat{k} \cdot \hat{z}$ and $f \equiv {\rm d ln}D/{\rm d ln}a$ is the linear growth rate with the scale factor $a=1/(1+z)$. Note that $k^2 = k^2_\perp + k^2_\parallel$, with $k_\parallel = \mu k$.

The Fisher matrix for a set of parameters $\{p\}$ is then given by \cite{Tegmark:1997rp}
\be
F_{\rm ij}=\frac{1}{8\pi^2}\int^1_{-1} d\mu \int^{k_{\rm max}}_{k_{\rm min}} k^2dk \; [\partial_i {\rm ln}P^{\rm S} \partial_j {\rm ln}P^{\rm S} ] V_{\rm eff} \, ,
\label{eq:Fish1}
\ee where $\partial_i \equiv \partial / \partial p_i$, and 
\be
V_{\rm eff} = V_{\rm sur} \left(\frac{P^{\rm S}}{P^{\rm S}+P^{\rm N}}\right)^2 \, ,
\ee with $V_{\rm sur}$ the survey (bin) volume and $P^{\rm N}$ the noise power spectrum. In general, we are going to work with multiple (independent) redshift bins of width $\Delta z = 0.1$, which means that the total Fisher matrix for each experiment is the sum of the Fisher matrices corresponding to each redshift bin. We will restrict our analysis to linear scales, imposing a non-linear cutoff at $k_{\rm NL} \simeq 0.1 \, (1+z)^{2/3} \, {\rm Mpc}^{-1}$ \cite{Smith:2002dz}, and thus ignore small-scale velocity dispersion effects.

\begin{figure}[H]
\centering
\includegraphics[width=\columnwidth]{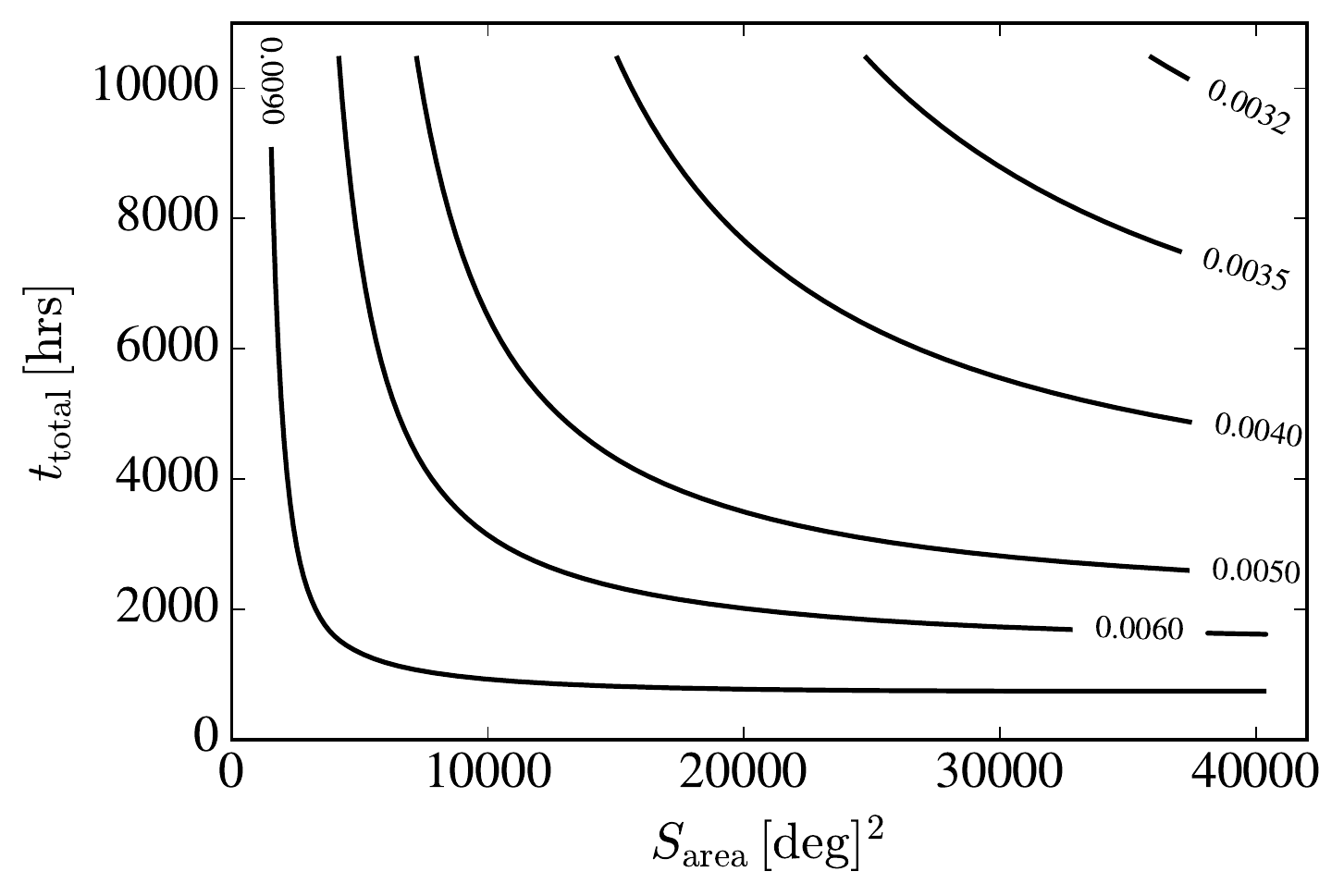}
\caption{Optimisation plot for SKA1-MID operating in single dish mode. We vary the survey strategy, i.e. the sky area $S_{\rm area}$ and the total observing time $t_{\rm total}$, calculating the $1-\sigma$ uncertainty on the spectral running $\alpha_{\rm s}$ keeping all other parameters fixed to their fiducial values. }
\label{fig:sd-opt}
\end{figure}

Having the Fisher Matrix formalism at hand, we would like to perform an optimisation study with respect to the survey strategy parameters, i.e. the sky area $S_{\rm area}$ and the total observing time $t_{\rm total}$. For this purpose we take the SKA1-MID array configuration and we calculate the $1-\sigma$ uncertainty on the spectral running $\alpha_{\rm s}$ (keeping all other parameters fixed to their fiducial values) at $k_{\star}=0.05 \, {\rm Mpc}^{-1}$. The results are plotted in Fig.~\ref{fig:sd-opt}. The forecasted uncertainty on $\alpha_{\rm s}$ decreases with increasing sky area and total observation time, but we notice that the contours have turning points beyond which they start to flatten. That is because of the non-trivial effect of the sky area to the total power spectrum measurement error
$
\delta P^{\rm HI} \propto \frac{1}{\sqrt {V_{\rm sur}}} \left( P^{\rm HI} + P^{N}\right) \, .
$
The cosmic variance error contribution decreases as $1/\sqrt{S_{\rm area}}$, but the contribution due to the thermal noise increases as $\sqrt{S_{\rm area}}$ (and decreases as $t_{\rm total}$). This means that there is a ``sweet spot" of a minimum sky area to achieve a certain precision for a given observing time. 

We are now going to calculate the forecasted uncertainties on $(n_{\rm s},\alpha_{\rm s},\beta_{\rm s})$ considering a large sky IM survey with SKA1-MID, operating in single-dish mode. This could be an $f_{\rm sky} \sim 0.7$ survey performed in $\sim 5,000 \, {\rm hrs}$. 
We find that such a survey would achieve 
$\sigma(n_{\rm s})=0.0022$, $\sigma(\alpha_{\rm s})=0.0043$, and $\sigma(\beta_{\rm s})=0.015$. Increasing the observing time to $\sim 10,000 \, {\rm hrs}$ (this is not an unrealistic scenario if the IM survey is commensal with other surveys), we get 
$\sigma(n_{\rm s})=0.0019$, $\sigma(\alpha_{\rm s})=0.0036$, and $\sigma(\beta_{\rm s})=0.013$.

Next we are going to consider a dedicated SKA2-MID-like experiment with $f_{\rm sky} = 0.7$ and thermal noise an order of magnitude lower than the first SKA1 case, which we achieve by increasing $t_{\rm total}$ by a factor of 2 and $N_{\rm dishes}$ (or $N_{\rm dishes} \times N_{\rm beams}$) by a factor of 5.
We find $\sigma(n_{\rm s})=0.0015$, $\sigma(\alpha_{\rm s})=0.0029$, and $\sigma(\beta_{\rm s})=0.009$. We could consider a configuration with even more dishes and/or feeds and the uncertainties would shrink even more, but this would be a very futuristic scenario; the above constraints can only probe significant deviations from the slow-roll single-field scenario --- we again note that in the usual single-field inflationary models the first running $|\alpha_{\rm s}| \sim 0.001$ \cite{Kosowsky:1995aa}, but models that produce a large running at the related wavenumber range also exist (see, for example, \cite{Silverstein:2008sg,Minor:2014xla}). 

\begin{figure}[H]
\centering
\includegraphics[width=\columnwidth]{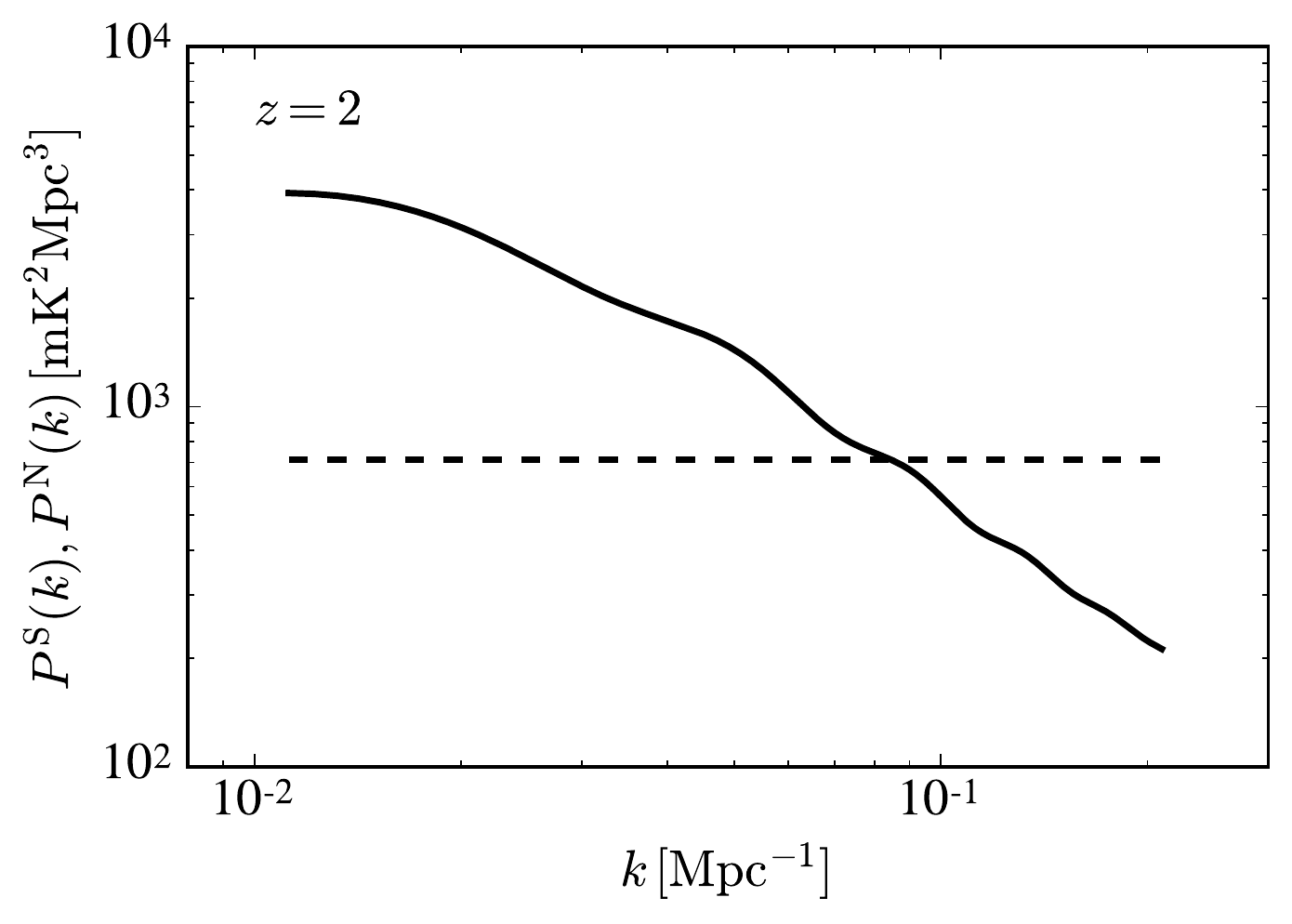}
\caption{The HI (solid) and thermal noise (dashed) power spectra at $z=2$ for a HIRAX-like survey (see main text for details).}
\label{fig:int-PN}
\end{figure}

Let us now move on to radio telescope arrays operating in the traditional interferometric mode. This mode is preferable for large scale cosmological studies in higher redshifts. That is because the largest scales the instrument can probe are determined by its FOV (not the sky area, like in single dish mode), hence at low redshifts the linear scales of interest are not accessible. The interferometer resolution is determined by the maximum baseline $D_{\rm max}$, which allows probing small scales. Note that the noise power spectrum of a radio interferometer with uniformly distributed antennae is flat. In our forecasts
we set $k_{\rm max}$ by comparing $k_{\rm NL}$ and $k^{\rm (int)}_{\rm max} \sim 2\pi D_{\rm max}/(r\lambda)$ at each redshift and choosing the one that is smaller. In Figure~\ref{fig:int-PN} we plot the HI signal (ignoring RSDs) and noise power spectra at $z=2$ for a HIRAX-like survey.

\begin{center}
\begin{table*}
\begin{tabular}{|l | c | c | c | c | c | c | c |}
\hline
IM survey & $k_{\star} \, [{\rm Mpc}]^{-1}$ & $f_{\rm sky}$ & $t_{\rm total}  \, [{\rm hrs}]$ & Redshift Range  & $\sigma\left(n_{\rm s}\right)$ &  $\sigma\left(\alpha_{\rm s}\right)$&  $\sigma\left(\beta_{\rm s}\right)$   \\             
\hline
\hline
SKA1-MID (sd) & $0.05$ & 0.70 & 5,000 & $0.35 < z < 3.05$  & 0.0022  & 0.0043 &  0.015  \\
SKA1-MID (sd) & \dittostraight& \dittostraight & 10,000 & \dittostraight & 0.0019  & 0.0036 &  0.013  \\
SKA2-MID-like (sd) & \dittostraight & \dittostraight & \dittostraight & \dittostraight & 0.0015 & 0.0029  & 0.009  \\
CHIME & 0.1 & 0.6 & \dittostraight & $0.8 < z <  2.5$ &  0.0013  & 0.0047 & 0.036 \\
HIRAX & 0.05 & 0.36 & \dittostraight & \dittostraight & 0.0020 & 0.0035  &  0.011  \\ 
HIRAX $({\rm higher} \, k_{\rm NL})$ & \dittostraight & \dittostraight & \dittostraight & \dittostraight & 0.0014 & 0.0020  &  0.007  \\ 
SKA2-LOW-like (compact) &\dittostraight & 0.6 & \dittostraight & $3.0 < z <  5.0$ & 0.0007  & 0.0008  & 0.003   \\	
\hline	
\end{tabular}
\caption{$1-\sigma$ forecasts for the HI intensity mapping surveys we consider in this study. The $(n_{\rm s},\alpha_{\rm s})$ uncertainties correspond to fixed $\beta_{\rm s} = 0$.}
\label{tab:IM}
\end{table*}
\end{center}

The dependence of the error on the power spectrum measurement on $S_{\rm area}$ and $t_{\rm total}$ is the same as in the single-dish mode case. Another parameter that is very important here is the covering fraction of the array, which can be written as $f_{\rm cover} = N_f (D_{\rm dish}/D_{\rm max})^2$. It describes how ``filled'' the array is with antennae, hence it cannot exceed unity. The thermal noise of the array is $\propto 1/f^2_{\rm cover}$, so there is a large difference depending on whether the array configuration is sparse or compact. This is the reason why purpose-build IM interferometers like CHIME and HIRAX are compact.

We can now forecast the constraints CHIME and HIRAX can give on the spectral index and its runnings. For CHIME, we need to change the pivot scale where the spectral index and its runnings are defined. That is because the scale $k = 0.05 \, {\rm Mpc}^{-1}$ is not accessible until $z \sim 1.6$. We are therefore going to use $k_{\star} = 0.1 \, {\rm Mpc}^{-1}$ 
(note that the reason that $k_{\star}$ is chosen to be $0.05 \, {\rm Mpc}^{-1}$ for Planck is because around that scale the tilt $n_{\rm s}$ and the first running $\alpha_{\rm s}$ decorrelate, so the constraint on $n_{\rm s}$ is optimal \cite{Cortes:2007ak}).
We find $\sigma(n_{\rm s})=0.0013$, $\sigma(\alpha_{\rm s})=0.0047$, and $\sigma(\beta_{\rm s})=0.036$. Repeating the calculation for HIRAX (with the standard choice $k_{\star} = 0.05 \, {\rm Mpc}^{-1}$) we get 
$\sigma(n_{\rm s})=0.0020$, $\sigma(\alpha_{\rm s})=0.0035$, and $\sigma(\beta_{\rm s})=0.011$.  The access to smaller scales using the interferometer mode means that increasing the non-linear cutoff at $k_{\rm NL} \simeq 0.14 \, (1+z)^{2/3} \, {\rm Mpc}^{-1}$ the HIRAX constraints improve significantly: $\sigma(n_{\rm s})=0.0014$, $\sigma(\alpha_{\rm s})=0.0020$, and $\sigma(\beta_{\rm s})=0.0074$.  These constraints on the runnings are the best we have obtained so far and the reasons are the access to smaller scales across a wide redshift range, the compactness of the HIRAX array, and the fact that its small dishes also allow for large scales to be probed.

Finally, we derive constraints on the SKA2-LOW-like compact configuration we described in Section~\ref{sec:surveys}, with $k_{\rm NL} \simeq 0.14 \, (1+z)^{2/3} \, {\rm Mpc}^{-1}$. The results are $\sigma(n_{\rm s})=0.0007$, $\sigma(\alpha_{\rm s})=0.0008$, and $\sigma(\beta_{\rm s})=0.003$. Using a 21cm intensity mapping survey with a compact array at high redshifts we can constrain $\alpha_{\rm s}$ at the level required for probing single-field slow-roll inflation.

\subsection{Optical galaxies}

\begin{center}
\begin{table*}
\begin{tabular}{|l | c | c | c | c | c | c |}
\hline
Optical galaxy survey & $k_{\star} \, [{\rm Mpc}]^{-1}$ & $f_{\rm sky}$  & Redshift Range  & $\sigma\left(n_{\rm s}\right)$ &  $\sigma\left(\alpha_{\rm s}\right)$&  $\sigma\left(\beta_{\rm s}\right)$   \\             
\hline
\hline
Stage 4 (Euclid-like) & $0.05$ & 0.36  &  $0.65 < z <  2.1$ & 0.0020 & 0.0038  & 0.010      \\  	
Stage 4 (Euclid-like) & $0.1$ & 0.36  &  $0.65 < z <  2.1$ & 0.0014 & 0.0030  & 0.010      \\
\hline	
\end{tabular}
\caption{$1-\sigma$ forecasts for the optical galaxy survey we consider in this study. The $(n_{\rm s},\alpha_{\rm s})$ uncertainties correspond to fixed $\beta_{\rm s} = 0$. }
\label{tab:gal}
\end{table*}
\end{center}

The Fisher matrix for an optical spectroscopic galaxy survey like Euclid is given by Equation~(\ref{eq:Fish1}), with 
\be
P^{\rm S} \equiv P^{\rm gg}(k,z;\mu) = [b^2_{\rm g}+f\mu^2]P(k,z) \, ,
\ee and $P^{\rm N}$ the shot noise, 
\be
P^{\rm N} = \frac{1}{\bar{n}} \, ,
\ee with $\bar{n}$ the number density of galaxies in the redshift bin under consideration and $b_{\rm g}$ the galaxy bias, which is assumed to be linear and deterministic on large scales. As we did in the IM case, we will consider the bias known (measured) in our forecasts.

The general rule for galaxy surveys is that increasing the sky area (hence the volume) results in 
improved constraints as the cosmic variance error is decreased. Decreasing the shot noise contribution by increasing the number density of galaxies also improves the constraints, up to the limit where the shot noise becomes negligible (see \cite{Adshead:2010mc} for a nice demonstration of this). Note that for a fixed $S_{\rm area}$, higher redshifts probe larger volumes and smaller scales become linear. However, the shot noise increases with redshift.

Using the Fisher matrix formalism for galaxy surveys we derive constraints assuming a Stage 4 Euclid-like spectroscopic survey; they are summarised in Table~\ref{tab:gal}. We find 
$\sigma(n_{\rm s})=0.0020$, $\sigma(\alpha_{\rm s})=0.0038$, and $\sigma(\beta_{\rm s})=0.010$ for $k_{\star}=0.05 \, {\rm Mpc}^{-1}$. At another pivot scale $k_{\star} = 0.1 \, {\rm Mpc}^{-1}$ $n_{\rm s}$ and $\alpha_{\rm s}$ become less correlated and their uncertainties are smaller:  $\sigma(n_{\rm s})=0.0014$, $\sigma(\alpha_{\rm s})=0.0030$. While this is useful for optimising the performance of a given survey, we will not explore it further in this work. Since we wish to combine the LSS forecasts with the ones from the CMB, we use $k_{\star} = 0.05 \, {\rm Mpc}^{-1}$. 

\begin{center}
\begin{table*}
\begin{tabular}{|l | c | c | c | c |}
\hline
Survey & $\sigma (n_{\rm s})$  &  $\sigma\left(\alpha_{\rm s}\right)$&  $\sigma\left(\beta_{\rm s}\right)$   \\             
\hline
\hline
Planck & 0.006 &  0.007 &    \\
COrE-like & 0.0019 &  0.0025 &  0.0058  \\
COrE-like + SKA1-MID (sd) & 0.0013 &  0.0021 &  0.0045  \\
COrE-like + SKA2-MID-like (sd) & 0.0011 & 0.0019  & 0.0042  \\
COrE-like + HIRAX & 0.0012 & 0.0020  &  0.0040  \\ 
COrE-like + HIRAX $({\rm higher} \, k_{\rm NL})$ & 0.0011 & 0.0015  &  0.0030  \\ 
COrE-like + SKA2-LOW-like (compact) & 0.0006 & 0.0007  & 0.0017   \\	
COrE-like + Euclid-like & 0.0011 & 0.0018  & 0.0037   \\	
\hline	
\end{tabular}
\caption{$1-\sigma$ forecasts for various surveys and combinations. The $(n_{\rm s},\alpha_{\rm s})$ uncertainties correspond to fixed $\beta_{\rm s} = 0$.}
\label{tab:comb}
\end{table*}
\end{center}

\subsection{ Combined forecasts }

We are now ready to combine our forecasts -- by adding the Fisher matrices -- using the COrE-like CMB survey and the various LSS surveys we considered in this study; the results are shown in Table~\ref{tab:comb}. Note that we only show the cases where $k_{\star} = 0.05 \, {\rm Mpc}^{-1}$ for the LSS survey, as this is the pivot scale chosen for the CMB measurements. For the SKA1-MID (sd) survey we use the $t_{\rm total} = 10,000 \, {\rm hrs}$ case. We also show the Planck constraints on ($n_{\rm s},\alpha_{\rm s}$) and the COrE-like forecasts for reference.

As expected, we find that combining surveys we get smaller uncertainties than in individual cases. The effect is more substantial for the IM surveys with compact interferometers targeting high redshifts and smaller scales, namely HIRAX and SKA2-LOW-like, and for the Euclid-like spectroscopic galaxy survey.

We should also note that if our main goal is to test single field inflation, we must concentrate on the first running $\alpha_{\rm s}$ as a $\beta_{\rm s} \sim 10^{-5}$ measurement is out of reach for the range of surveys we have considered.

\section{Conclusions}
\label{sec:conclusions}

In this work we have investigated the prospects of utilising future datasets from 21cm intensity mapping, CMB, and optical galaxy surveys, in order to constrain the primordial Universe. The purpose of our study was two-fold: we wanted to assess the possibility of using the innovative 21cm intensity mapping technique to probe and constrain the scalar spectral index and its runnings with ongoing and future experiments. We also wanted to demonstrate how the synergies between a future CMB survey and large scale structure surveys can improve the results from the former alone.

We should comment on various assumptions and simplifications we made in this study. A major concern for intensity mapping surveys is the foreground contamination problem from galactic and extragalactic sources. These can be orders of magnitude larger than the signal, but if data calibration is done properly we can use their spectral smoothness to remove them (see, for example, \cite{Wolz:2015sqa, Alonso15, Olivari:2015tka}). Another way to mitigate the foreground problem is cross-correlating the 21cm intensity maps with optical galaxies, which can also help with alleviating systematic effects that are relevant to one type of survey but not the other \cite{Masui:2012zc, Pourtsidou:2015mia}. These ideas and methods can be tested in the near future with IM pathfinder surveys using instruments like BINGO \cite{Battye:2012tg} and MeerKAT\footnote{www.ska.ac.za/science-engineering/meerkat}. Note that the possibility of investigating another inflationary feature, namely primordial non-Gaussianity, with intensity mapping and optical galaxy surveys has been investigated recently in \cite{Fonseca:2016xvi}.

Another way to constrain the spectral index and its runnings using the redshifted 21cm radiation is by probing the Epoch of Reionization and the Dark Ages (see, for example, \cite{Mao:2008ug, Barger:2008ii, Adshead:2010mc, Munoz:2016owz}). In \cite{Munoz:2016owz} it was found that an interferometer similar to the proposed Fast Fourier Transform Telescope \cite{Tegmark:2008au}  with a $1 \, {\rm km}$ baseline could achieve $\delta \alpha_{\rm s} = 0.001$, while a -- very futuristic -- lunar interferometer targeting the dark ages could reach $\delta \alpha_{\rm s} \sim 10^{-5}$. In the same study predictions were also made for various optical and HI galaxy surveys (performed with instruments like WFIRST\footnote{https://wfirst.gsfc.nasa.gov}, DESI\footnote{http://desi.lbl.gov}, and the SKA), and the results are in general comparable to ours. Traditional galaxy surveys (either in the optical or the radio) and intensity mapping have different strengths and weaknesses, but we believe it is imperative to exploit the possible synergies between them, in order to get more precise and robust cosmological measurements. For this study in particular, the fact that IM experiments can easily give us access to high redshifts means that we can use a vaster range of linear scales. However, it would be extremely beneficial, for both galaxy and IM experiments, if we could use some of the non-linear scales information. Here we chose to work with conservative non-linear cutoffs, but if we could accurately model non-linearities the leverage would be immense and these surveys would directly compete with the best CMB experiments --- this would not affect the constraints on the scalar spectral index that much, but it would greatly improve the measurements of $\alpha_{\rm s}$, which is what we are mainly after \cite{Takada:2005si, Adshead:2010mc}.  

To conclude, we have shown that future CMB, 21cm intensity mapping, and optical galaxy surveys can be used to improve our knowledge of the primordial Universe and constrain the extensive model space of single-field, slow-roll inflation. We believe that the results of our study provide strong motivation for maximising the synergistic power of future CMB and multi-wavelength large scale structure surveys.

\section{Acknowledgments}
I acknowledge support by a Dennis Sciama Fellowship at the University of Portsmouth. I acknowledge use of the \texttt{CAMB} code \cite{camb}. I would like to thank Robert Crittenden and Vincent Vennin for useful discussions and comments on the manuscript. 
Fisher matrix codes used in this work are available from \url{https://github.com/Alkistis/Inflation}.

\bibliographystyle{apsrev}
\bibliography{runnings}

\end{document}